\documentclass[prl,twocolumn,showpacs]{revtex4}
\usepackage{graphicx}
\begin{document}
\title{Reversal of the Charge Transfer between Host and Dopant Atoms
in Semiconductor Nanocrystals} 
\author{Torbj\"orn Blomquist}
\email{tblomqui@sfu.ca} \author{George Kirczenow}
\email{kirczeno@sfu.ca} \affiliation{ Department of Physics, Simon
Fraser University, Burnaby, British Columbia, Canada V5A 1S6}

\begin{abstract}
We present {\em ab initio} density functional calculations that show P
(Al) dopant atoms in small hydrogen-terminated Si crystals to be
negatively (positively) charged. These signs of the dopant charges are
{\em reversed} relative to the same dopants in bulk Si. We predict
this novel reversal of the dopant charge (and electronic character of
the doping) to occur at crystal sizes of order 100 Si atoms. We
explain it as a result of competition between fundamental principles
governing charge transfer in bulk semiconductors and molecules and
predict it to occur in nanocrystals of most semiconductors.
\end{abstract}

\date{\today}
\maketitle

Introducing appropriate impurity atoms (known as ``dopants") into a
semiconductor can dramatically affect electrical conduction in the
material and is key to the operation of modern electronic
devices.\cite{Kittel} The dopant atoms modify the conductivity of the
semiconductor by supplying it with additional free electrons or holes
that can carry an electric current.  If an impurity atom having one
more electron than an atom of the semiconductor host replaces a host
atom, in many cases the extra electron is very weakly bound to the
impurity atom in the solid state environment.\cite{Kohn} Thus at room
temperature the shallow impurity loses ({\em donates}) the extra electron to
the semiconductor and the impurity atom becomes positively
charged. Conversely an impurity atom with one fewer electron than the
host {\em accepts} an electron from the host, becomes negatively
charged, and a positively charged free hole appears in the
semiconductor. This qualitative picture of charge transfer between
semiconductor host and shallow dopant is well established for bulk
semiconductor materials and is fundamental to our understanding of the
properties of semiconductor devices. Recent experimental and
theoretical work \cite{Mimura, Melnikov, Pandey, Zhu, Lannoo,
Einevoll, Stebe,Estreicher} has shown that it also holds for a variety
of doped semiconductor nanoparticles.

\begin{figure}[Htb] 
\includegraphics[width=0.45\textwidth]{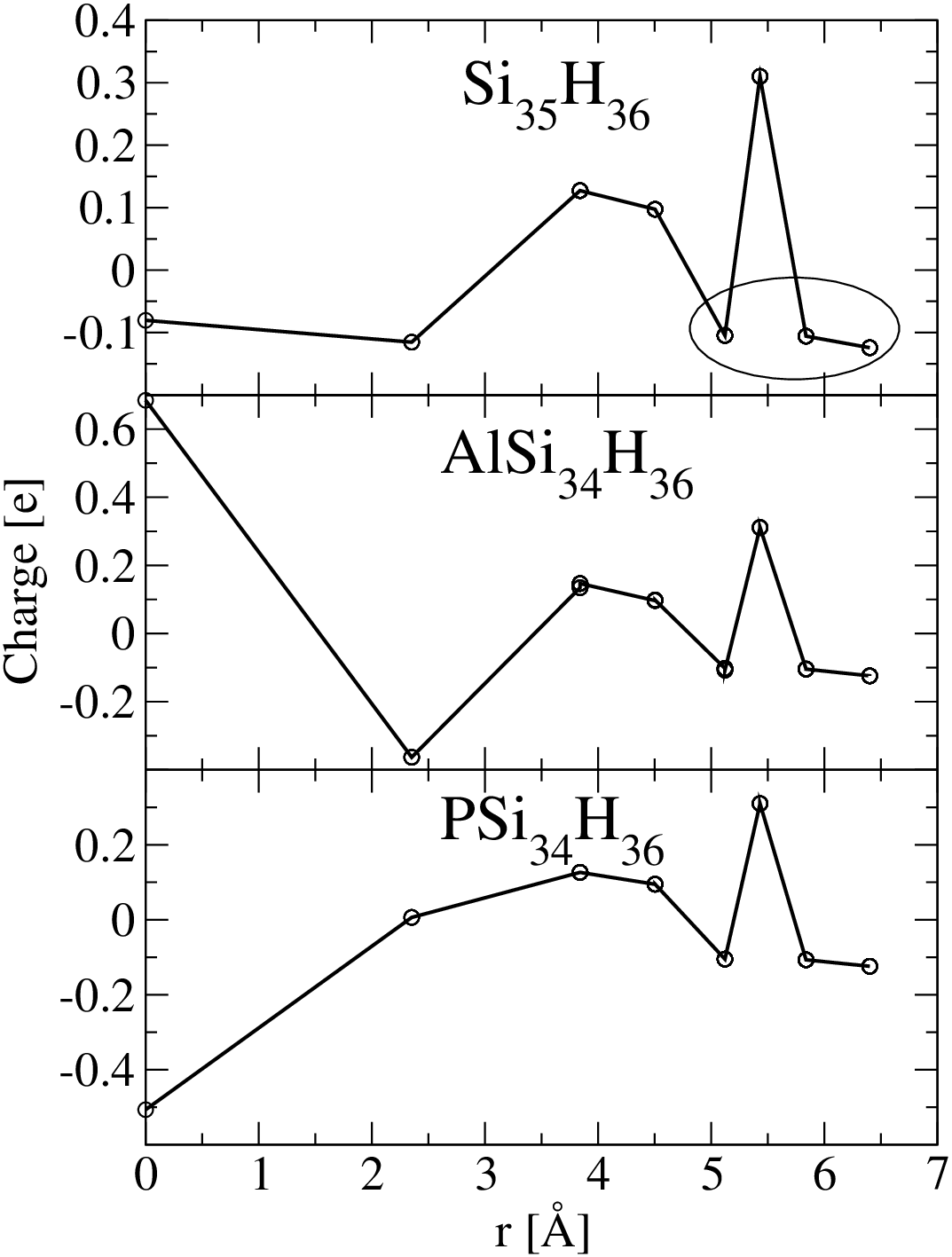}
\caption{\label{gaussianfig}On site natural population \cite{nat}
charge as function of the site radial coordinate for
Si$_{35}$H$_{36}$. Undoped crystallite in top graph. The ellipse in the
top graph shows which atoms are hydrogen. The central atom (at $r=0$)
has been replaced with aluminum (acceptor in bulk Si) in center graph
and with phosphorus (donor in bulk Si) in bottom graph.}
\end{figure}

Charge transfer also plays an important role in the chemistry of
molecular systems \cite{Stoker}, however, the basic principles that
apply in this case are different: Atoms are classified according to
their {\em electronegativity} which is defined so that an atom with a
larger electronegativity will attract (negative) electronic charge
from an atom with smaller electronegativity.  Atoms with nearly filled
valence orbitals have large electronegativities because filled
orbitals are energetically stable. Conversely atoms with nearly empty
valence orbitals have low electronegativities. Simple semiconductors
consisting of atoms from group IV of the periodic table have precisely
half-filled valence orbitals. A group V atom has one more valence
electron and since its valence orbitals are closer to being filled it
has a larger electronegativity. Therefore, according to this picture,
a dopant atom with one more electron than the host semiconductor
should attract charge from the surrounding host atoms and carry a
negative charge. Similarly a group III dopant atom with one fewer
electron than the host would be positive. Clearly this reasoning based
on considerations of molecular chemistry predicts charge transfer
between dopant and host {\em opposite in sign} to that found in the
solid state semiconductor systems discussed above.  This raises the
intriguing, and until now unrecognized, possibility that charge
transfer in doped semiconductor nanocrystals with dimensions
approaching the molecular scale might differ fundamentally from that
in macroscopic semiconductors, with profound implications for the
electronic properties of nanoscale semiconductor devices. Here we show
theoretically that this is in fact the case and explore the crossover
between the conventional (macroscopic) and novel
nanoscopic/quasi-molecular (reversed charge transfer) doping regimes.

Representative results of our {\em ab initio} density functional
calculations\cite{Gaussian} of the charge distributions in some small
H-terminated Si crystals doped with P and Al are shown in Fig.
\ref{gaussianfig}. Since a P (Al) atom has one more (less) valence
electron than Si, in bulk Si the shallow dopant P (Al) is an electron
donor (acceptor) and the impurity site is positively (negatively)
charged. In Fig. \ref{gaussianfig}, however, the reverse is true. Thus
our results for these Si nanocrystals clearly demonstrate the
existence of the quasi-molecular regime that we have proposed above,
where charge transfer is governed by electronegativity considerations
rather than by the standard theory of doping in bulk semiconductors.

{\em Ab initio} quantum chemistry calculations such as those that
yielded the results of Fig. \ref{gaussianfig} cannot at present be
made for much larger crystallites due to practical limitations of
computers. We have therefore developed a Poisson-Schr\"odinger (PS)
model for silicon based on a nonorthogonal tight-binding (TB) model in
order to explore the crossover from the quasi-molecular regime to
crystals large enough that standard semiconductor theory should be
appropriate. We have also examined how energy gaps and dopant levels
evolve with the size of the Si crystallite.

We have based our TB model on that of Bernstein et al.\cite{Bernstein}
which reproduces the band structure of silicon very well, and gives
reasonable values for electron and hole masses, see table
\ref{masstable}. The on-site potentials in the Bernstein model are
however functions of the local density of atoms but differences in
on-site potentials are explicitly given by the PS scheme, so we have
used Bernstein's values for bulk Si as a starting point for the PS
scheme. On-site potentials for hydrogen and hopping integrals for Si-H
have been fitted to reproduce charge distributions obtained from {\em
ab initio} density functional calculations.\cite{Gaussian} We have
used the same overlap and hopping integrals for Al-Si and P-Si as for
Si-Si. The on-site parameters for Al and P have been fitted to yield
the correct sign of the charge on the impurity site for small crystals
and realistic values for dopant energy levels for large ones.  The
on-site electron repulsion energies for Al, Si and P are based on
valence orbital ionization energies taken from table D4 of
Ref. \onlinecite{McGlynn}; equation (D6) in Ref. \onlinecite{McGlynn}
is used for H. We have ensured that our model reproduces the on-site
energies and band structure of the Bernstein model for bulk Si. The
energy on site $i$ includes the electrostatic term
\begin{equation}
V_i=\frac{1}{\epsilon_r} \sum_{j\neq i} 
\frac{q_j}{|\mathbf{r}_i-\mathbf{r}_j|}+
\int_S\frac{\rho(\mathbf{r})}{|\mathbf{r}-\mathbf{r}_i|}d\mathbf{r},
\end{equation}
where $\epsilon_r$ is the relative dielectric constant due to core
polarization(core electrons are not included in the TB model) in Si,
$q_j$ is the net charge on site $j$, $\mathbf{r}_{i}$ is the position
of site $i$, $S$ is the surface of the structure and
$\rho(\mathbf{r})$ is the surface polarization charge, due to the core
polarization. We have chosen the value $\epsilon_r=6.5$ so that the
model reproduces the correct total dielectric constant of 11.8 for
infinite Si slabs, taking into account polarization of both the
valence and core electrons.  The electrostatic equation is solved
self-consistently together with the Schr\"odinger equation
$H\psi=ES\psi$ for the nonorthogonal TB model. Here $S$ are overlap
integrals.  The charge on each site is calculated using Mulliken
population analysis.\cite{McGlynn}

\begin{table}
\caption{\label{masstable}Some properties of the TB
model,\cite{Bernstein} experimental values are given in
parenthesis.\cite{Madelung}}
\begin{ruledtabular}
\begin{tabular}{lll}
Position of conduction band minima & 87.7\% $\Gamma-$X &(85\%)\\
Band gap & 1.01{\thinspace}eV & (1.12)\\ 
Light hole mass & 0.26{\thinspace}$m_e$ & (0.15)\\ 
Heavy hole mass & 0.31{\thinspace}$m_e$ & (0.54)\\ 
Longitudinal electron mass & 0.55{\thinspace}$m_e$ & (0.92)\\
Transverse electron mass & 0.15{\thinspace}$m_e$ & (0.19)\\
\end{tabular}
\end{ruledtabular}
\end{table}

\begin{figure}[Htb] 
\includegraphics[width=0.45\textwidth]{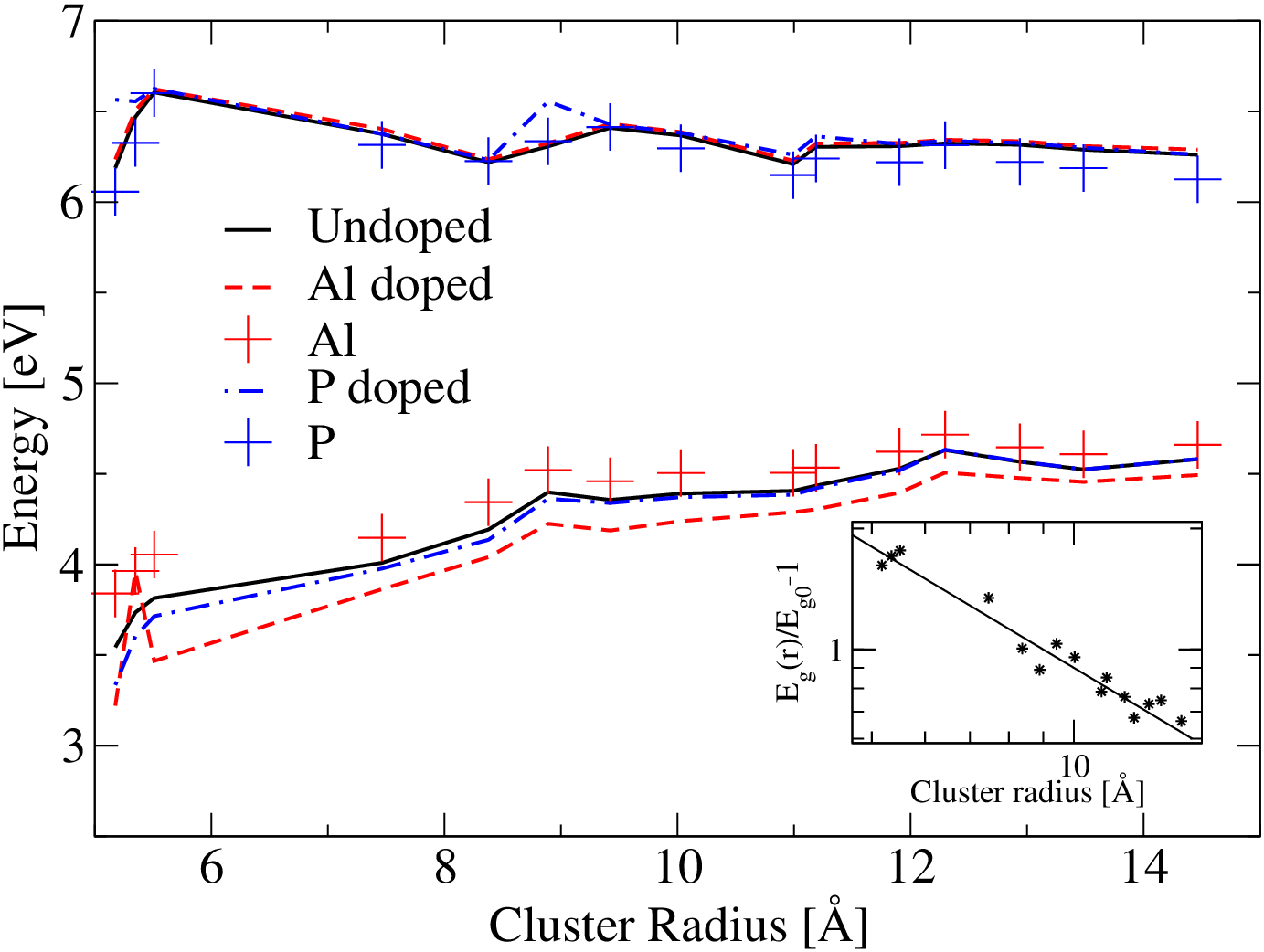}
\caption{\label{bandgapfig}(Color online) The energies of the valence
and conduction bands of the crystallites, with and without dopants,
plotted as functions of crystallite radius. The energies of the dopant
states are also shown. The inset shows the band gap plotted as
$f(r)=E_g(r)/E_{g0}-1$ on a log-log-scale for comparison with papers
\onlinecite{Liu,Zunger}. A fit,$f(r)=9.0r^{-1.0}$ is also shown.}
\end{figure}

\begin{table}
\caption{\label{energiestable} Band gaps for undoped and doped
crystallites. Dopant levels for the doped nanocrystals.}
\begin{ruledtabular}
\begin{tabular}{lrrrrr}
Crystallite & gap & P gap & Al gap &
P level& Al level\\ 
& (eV) & (eV) & (eV) & (meV) & (meV)\\
\hline
Si$_{29}$H$_{36}$   & 2.64567 & 3.22941 & 3.01460 & 507.92 &  618.78 \\
Si$_{32}$H$_{36}$   & 2.73479 & 2.95265 & 2.54536 & 229.18 &    2.46 \\
Si$_{35}$H$_{36}$   & 2.79060 & 2.91065 & 3.15534 &  24.37 &  587.29 \\
Si$_{87}$H$_{76}$   & 2.36659 & 2.39807 & 2.53979 &  60.15 &  284.22 \\
Si$_{123}$H$_{100}$ & 2.02584 & 2.09530 & 2.19588 &   5.69 &  303.12 \\
Si$_{147}$H$_{100}$ & 1.90885 & 2.19233 & 2.09923 & 219.87 &  295.63 \\
Si$_{175}$H$_{116}$ & 2.05314 & 2.08903 & 2.24444 &  14.17 &  271.98 \\
Si$_{211}$H$_{140}$ & 1.97595 & 2.00978 & 2.14665 &  84.48 &  265.58 \\
Si$_{278}$H$_{172}$ & 1.80325 & 1.87565 & 1.93952 & 111.71 &  217.66 \\
Si$_{293}$H$_{172}$ & 1.86942 & 1.94209 & 2.01869 & 122.92 &  229.42 \\
Si$_{353}$H$_{196}$ & 1.77987 & 1.80023 & 1.93001 & 100.28 &  226.42 \\
Si$_{389}$H$_{196}$ & 1.69254 & 1.70375 & 1.83677 &  22.85 &  209.68 \\
Si$_{453}$H$_{228}$ & 1.74891 & 1.75739 & 1.85943 & 105.10 &  169.72 \\
Si$_{513}$H$_{252}$ & 1.76443 & 1.77445 & 1.85331 & 111.99 &  152.69 \\
Si$_{633}$H$_{300}$ & 1.68004 & 1.67888 & 1.79626 & 135.37 &  166.78 \\
\end{tabular}
\end{ruledtabular}
\end{table}

We have applied the present model to calculate the ground state
properties of a number of silicon nanocrystals ranging in size from
from Si$_{29}$H$_{36}$ to Si$_{633}$H$_{300}$, with and without
dopants.  All of them are approximately spherical, have tetrahedral
symmetry, and are hydrogen terminated to obtain a clean energy gap for
the undoped nanocrystals.\cite{Liu}

Figure \ref{bandgapfig} and Table \ref{energiestable} show how the
energies of the valence band, conduction band and dopant levels change
with nanocrystal size. The conduction band energy varies little with
nanocrystal size and dopant species. The valence band moves up
narrowing the band gap to $\sim$1.7{\thinspace}eV for
Si$_{633}$H$_{300}$ from $\sim$2.7{\thinspace}eV for
Si$_{29}$H$_{36}$. Doping widens the band gap somewhat, but this
effect is most significant for the smallest crystals. The band gap can
be fitted to a function ${E_g(d)}/{E_{g0}}-1=Ar^{-b}$, where
$r=1.68456N^{1/3}$ is the radius of the crystallite, $N$ is the number
of Si atoms, $E_{g0}$ is the band gap in the bulk and, $A$ and $b$ are
fitting parameters. We find that $A=9.0$ and $b=1.0$; see fitted line
in inset of Fig. \ref{bandgapfig}. Liu et al.\cite{Liu} and Zunger et
al.\cite{Zunger} report $b=1.37$ in models without Coulomb
interactions.  Effective mass theory (particle in a box) predicts an
$r^{-2}$ scaling. Our result, however, agrees very well with density
functional theory calculations \cite{Delley,Melnikov,Ogut}.

\begin{figure}[Htb] 
\includegraphics[width=0.45\textwidth]{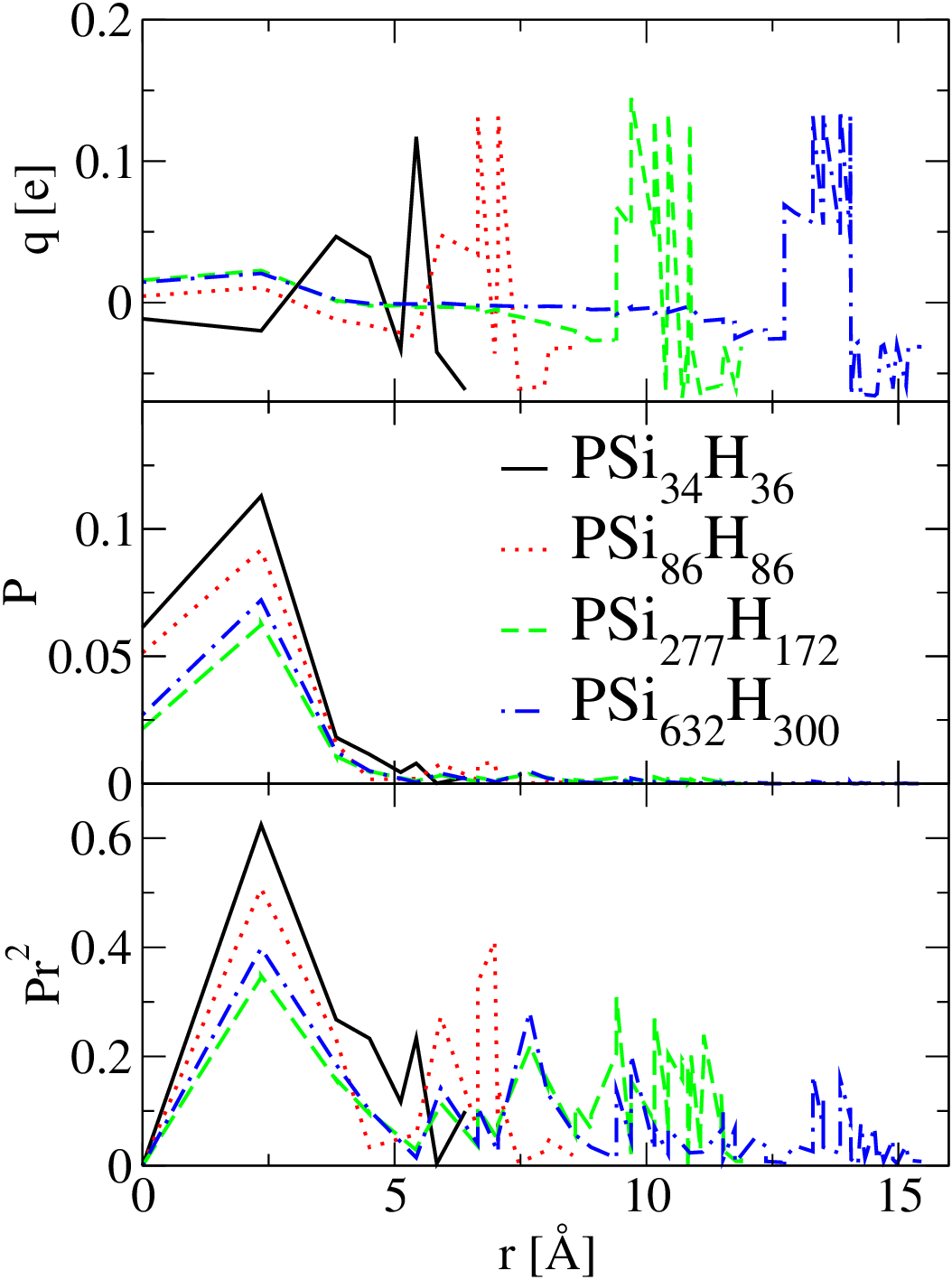}
\caption{\label{DopingstatePfig}(Color online) Top graph, total
Mulliken charge on each site as function of radial coordinate for four
different P-doped clusters.  Center graph, Mulliken probability
distribution for the donor state. Bottom graph, probability
distribution multiplied by radial coordinate squared to approximate
probability of finding the electron at a certain radius.}
\end{figure}

\begin{figure}[Htb] 
\includegraphics[width=0.45\textwidth]{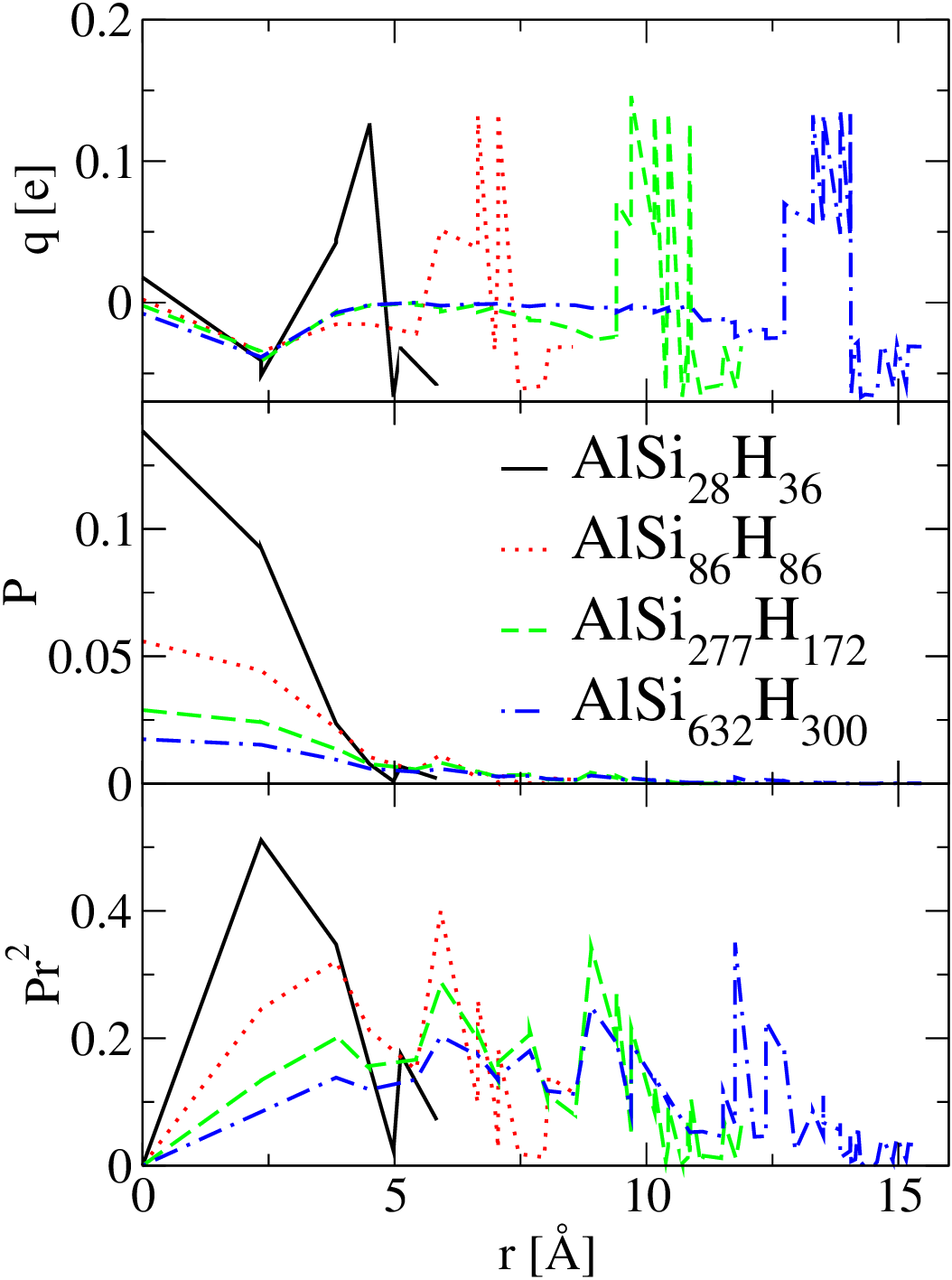}
\caption{\label{DopingstateAlfig}(Color online) Top graph, total
Mulliken charge on each site as function of radial coordinate for four
different Al-doped clusters.  Center graph, Mulliken probability
distribution for the acceptor state. Bottom graph, probability
distribution multiplied by radial coordinate squared to approximate
probability of finding the hole at a certain radius.}
\end{figure}

There has been interest in how the system size affects the dopant
levels and there has been a number of studies using different methods,
for example effective mass theory,\cite{Zhu,Pandey,Stebe}
TB,\cite{Einevoll} PRDDO\cite{Estreicher} and DFT.\cite{Lannoo} 
Our results are consistent with this previous work. We
define the dopant level as the energy difference between the partly
filled dopant state and its nearest neighbor state.  The Al dopant
levels in the nanocrystals, see Table \ref{energiestable}, vary quite
smoothly with crystal size (AlSi$_{31}$H$_{36}$ is an exception), down
to about three times the bulk value (57{\thinspace}meV) for
AlSi$_{632}$H$_{300}$. The P dopant levels on the other hand vary a
lot from cluster to cluster, but also reach about three times the bulk
value (45{\thinspace}meV) for our largest crystallite. We attribute
this difference between the acceptor and donor states to the fact that
the electron states have a larger probability on the surface sites,
making donor states much more sensitive to the surface than acceptor
states which typically are more localized to the interior of the
cluster; see the bottom graphs of Figs. \ref{DopingstatePfig} and
\ref{DopingstateAlfig}. The strong variations for the donor state
levels suggest that it would be difficult to engineer the properties
of a small n-doped cluster without atomic control in the
manufacturing. It is also relevant in this regard that the bands in
these small structures are made up from discrete energy levels and
even for our largest crystallites, these levels have an energy spacing
of $5-50${\thinspace}meV.

The charge on the impurity site ($r=0$) for the phosphorus doped
crystallites in Fig. \ref{DopingstatePfig} exhibits a crossover from
negative to positive when going from small to large nanocrystals.
This crossover between the quasi-molecular behavior and the bulk
semiconductor behavior occurs between PSi$_{34}$H$_{36}$ and
PSi$_{86}$H$_{76}$. For the aluminum doped crystallites
(Fig. \ref{DopingstateAlfig}) we find a crossover from a positive to a
negative impurity site between AlSi$_{122}$H$_{100}$ and
AlSi$_{146}$H$_{100}$. The precise crossover points are sensitive to
the parameters of the model and this result should be regarded as the
first (order of magnitude) estimate.  The charges on the impurity site
in Figs. \ref{DopingstatePfig} and \ref{DopingstateAlfig} are
consistently somewhat smaller in magnitude than in
Fig. \ref{gaussianfig}; we attribute this difference to the fact that
Mulliken population analysis (Figs. \ref{DopingstatePfig} and
\ref{DopingstateAlfig}), tends to smear charge between overlapping
orbitals on neighboring atoms more than do natural orbital
calculations (Fig. \ref{gaussianfig}).

In conclusion: The Poisson-Schr\"odinger model we have developed has
allowed us to explore the crossover from a novel regime in
semiconductor nanocrystals in which the molecular view of charge
transfer between atoms holds true to a regime where macroscopic solid
state semiconductor theory prevails.  The crossover is signaled by a
striking reversal of the sign of the charge transfer between the host
semiconductor and dopant atom that has not been anticipated in
previous experimental or theoretical work.  We predict that it should
occur at nanocrystal sizes of order 100 Si atoms.  Since very basic
principles of solid state semiconductor physics and molecular
chemistry are the underlying reasons for the charge reversal, we
predict it to be a general phenomenon occurring for a wide variety of
nanoscopic semiconductors and dopants. For Si nanocrystals we also
find an energy gap widening that scales as $r^{-1.0}$ consistent with
density functional theory calculations.\cite{{Delley,Melnikov,Ogut}}
We predict the dopant energy levels for Al in Si nanocrystals to vary
quite smoothly with cluster size while donor levels should vary widely
from crystallite to crystallite, making it difficult to engineer
properties of P-doped Si nanocrystals without atomic control in
manufacturing.

This work was supported by NSERC and the Canadian Institute for
Advanced Research.


\begin{thebibliography}{}

\bibitem{Kittel} C. Kittel, \textit{Introduction to solid state
physics} (Wiley, NewYork, 1996), 7th ed.

\bibitem{Kohn} W. Kohn, Phys. Rev. \textbf{105}, 509 (1957).

\bibitem{Mimura} A. Mimura, M. Fujii, S. Hayashi, D.
Kovalev and F. Koch, Phys. Rev. B \textbf{62} 12625 (2000).

\bibitem{Melnikov} D. V. Melnikov and J. R. Chelikowsky,
Phys. Rev. Let. \textbf{92} 046802 (2004).

\bibitem{Pandey} R. K. Pandey, M. K. Harbola and V.y A. Singh,
cond-mat/0308029 (2003).

\bibitem {Zhu} J.-L. Zhu, J.-J. Xiong and B.-L. Gu,
Phys. Rev. B \textbf{41}, 6001 (1990).

\bibitem{Stebe} B. St\'eb\'e, E. Assaid, F. Dujardin, and S. Le Goff,
Phys. Rev. B \textbf{54}, 17785 (1996).

\bibitem{Einevoll} G. T. Einevoll and Y.-C. Chang, Phys. Rev. B
\textbf{40}, 9683 (1989).

\bibitem{Estreicher} S. Estreicher, Phys. Rev. B \textbf{37} 858 (1988). 

\bibitem{Lannoo} M. Lannoo, C. Delerue, and G. Allan,
Phys. Rev. Lett. \textbf{74}, 3415 (1995).

\bibitem{Stoker} H. S. Stoker, \textit{Introduction to Chemical
Principles} (Macmillan, New York, 1993), 4th ed.

\bibitem{Gaussian} We employed the \textsc{gaussian 98} numerical
implementation of density functional theory with the 6-31G(d) basis
set and the B3LYP exchange-correlation energy functional.

\bibitem{Bernstein} N.Bernstein, M. J. Mehl, D. A.
Papaconstantopoulos, N. I. Papanicolaou, M Z. Bazant ,and
E. Kaxiras, Phys. Rev. B \textbf{62}, 4477 (2000),
ibid. \textbf{65}, 249902 (E) (2002).

\bibitem{McGlynn} S. P. McGlynn, L. G. Vanquickenborne, M. Kinoshita,
D. G. Carroll, \textit{Introduction to Applied Quantum Chemistry}
(Holt, Rinehart and Winston, Inc, New York, 1972)

\bibitem{Liu} L. Liu, C.S. Jayanthi, S.-Y. Wu, cond-mat/0012217
(2003).

\bibitem{Zunger} A. Zunger, L.-W. Wang,
Appl. Surf. Sci. \textbf{102}, 350 (1996).

\bibitem{Delley} B. Delley and E. F. Steigmeier, Phys. Rev. B
\textbf{47}, 1397 (1993); Appl. Phys. Lett. \textbf{67}, 2370 (1995).

\bibitem{Ogut} S. \"O\u g\"ut and J. R. Chelikowsky, and S. G. Louie,
Phys. Rev. Lett. \textbf{79}, 1770 (1997).

\bibitem{nat} A. E. Reed, L. A. Curtiss and F. Weinhold, Chem
Rev. \textbf{88}, 899 (1988).

\bibitem{Madelung} O. Madelung, \textit{Semiconductors Group IV
Elements and III-V Compounds} (Springer-Verlag, Berlin, 1991)

\end{thebibliography}
\end{document}